\documentclass[12pt,prd,draft,nofootinbib]{revtex4}

\usepackage{amsfonts}
\usepackage{amssymb}
\usepackage{amsmath}
\usepackage{bm}
\usepackage{latexsym}

\newcommand{\beq}{\begin{eqnarray}}
\newcommand{\eeq}{\end{eqnarray}}
\newcommand{\be}{\begin{equation}}
\newcommand{\ee}{\end{equation}}

\def\fun#1#2{\lower3.6pt\vbox{\baselineskip0pt\lineskip.9pt
\ialign{$\mathsurround=0pt#1\hfil ##\hfil$\crcr#2\crcr\sim\crcr}}}

\newcommand{{\SD}}{\rm SD}

\newcommand{\vep}{\bm p}

\begin{document}
\title{Radial Regge trajectories for higher $\psi(nS)$ and $\psi(nD)$ states}
\author{A.M.Badalian}
\affiliation{State Research Center, Institute of Theoretical and
Experimental Physics, Moscow, 117218 Russia}
\email[E-mail;]{badalian@itep.ru}
\date{\today}

\begin{abstract}
The masses of  $\psi((n+1)\,{}^3S_1)$ and $\psi(n\,{}^3D_1)$ are
calculated using the relativistic string Hamiltonian with
``linear+gluon-exchange" potential. They occur in the range
4.5--5.8~GeV, in particular, $M(3D)=4.54$~GeV, $M(5S)=4.79$~GeV,
$M(4D)=4.85$~GeV are calculated with accuracy $\sim 50$~MeV. For
higher charmonium states linear Regge trajectories:
$M^2(nS)=M^2(\psi(4.42))+ 2.91$~GeV$^2~(n-4)$ $(n\geq 4) $   and
$M^2(nD)=(4.54^2+ 2.88(n-3))$~GeV$^2$ ( $n\geq 3$) are obtained
only for higher charmonium states. They have a slope two times
larger than that of light mesons and  give a good description of
calculated masses. These masses are compared to enhancements in
some recent $e^+e^-$ experiments.
\end{abstract}

\maketitle

\section{Introduction}

Observation of higher charmonium states is  very important for
theory, first of all, to understand the $c\bar c$ dynamics at
large distances. At present only the $\psi(4415)$ resonance,
discovered long ago in 1976 \cite{ 1}, is well established; its
mass, $M(4415)=4421\pm 7$ MeV, is now known with a good accuracy
\cite{ 2,3}. However, even for this resonance there is an
uncertainty in the value of its dielectron width \cite{3}. The
analysis of most precise BES data on the ratio
$R=\sigma(e^+e^-\rightarrow$ hadrons)$~/\sigma(e^+e^-\rightarrow
\mu^+\mu^-)$ in \cite{4} has given
$\Gamma_{ee}(\psi(4415))=0.37\pm 0.14$~KeV, while in \cite{5} from
the same experimental $R$ values four different
$\Gamma_{ee}(\psi(4415))$, in the range 0.45--0.78~KeV, have been
extracted in different fits. Meanwhile precise knowledge of
dielectron widths of higher charmonium states may give an
important information on the $S-D$ mixing and different decays.

Therefore in our paper we concentrate on the masses for the higher
$nS$ and $nD$ charmonium states. Although the resonances, like
$\psi(3D)$, $\psi(5S)$, and $\psi(4D)$, are not well established
yet, several enhancements in the range 4.5--5.0~GeV were observed
in a number of recent $e^+e^-$ experiments: in $e^+e^-\rightarrow
D^0\bar D^{*-}\pi^+$, $D^{*+}D^{*-}$, $D_s^{*+}D_s^{*-}$ \cite{6},
$e^+e^-\rightarrow \Lambda_c^+\Lambda_c^-$ \cite{ 7} of the Belle
Collab., and also in the BaBar data on $e^+e^-\rightarrow
\pi^+\pi^-J/\psi$ \cite{ 8}, $e^+e^-\rightarrow D^*\bar D^*$,
$e^+e^-\rightarrow D\bar D^*$ via the initial state state
radiation \cite{9}. These enhancements have been analysed in
\cite{ 10,11}, where they are interpreted as the $\psi(3D)$,
$\psi(5S)$, and $\psi(4D)$ vector charmonium states, and their
masses and total widths were extracted from fits to experimental
data.

Here we consider only conventional $c\bar c$ mesons in the
framework of relativistic string picture. We perform calculations
of two kinds: with a universal linear + gluon-exchange (GE)
potential \cite{12} and also for purely linear potential when GE
interaction is taken as a perturbation. We shall show that linear
confining potential dominates in $c\bar c$ dynamics at large
distances, thus simplifying an analysis for several reasons.

Firstly,  at large distances GE potential is small as compared to
confining term. Its typical contribution  to the energy excitation
$E(nl)$ ($3\leq n\leq 8$) is of order 150~MeV, while a
contribution from linear potential is $\sim 1.5-2.2$~GeV.
Therefore the masses of higher states weakly depend on the
parameters of GE potential, which may be very much different even
in QCD motivated models \cite{ 13,14}.

Secondly, higher states have large sizes and their hyperfine and
fine-structure splittings are small, so that their masses
practically coincide with the centroid masses. Thus one escapes
uncertainties coming from parameters of spin-dependent potentials
\cite{15}.

Also we assume here that hadronic shifts of higher resonances due
to open channel(s) are not large, being of the same order as for
low-lying states, which are typically $\simeq 40$~MeV \cite{
16,17}, and only for $X(3872)$ a hadronic shift is larger, $\sim
70$~MeV, due to specifically strong coupling of the $P-$wave
charmonium state to the $S-$wave threshold. In this respect the
situation in charmonium differs from that of light mesons, where
hadronic shifts of radial excitations are large and a creation of
virtual quark-antiquark pairs should be taken into account \cite{
18}. Hence, we can perform calculations in single-channel
approximation, estimating an accuracy of our calculations as $\pm
50$~MeV.

We use here the relativistic string Hamiltonian (RSH) \cite{
19,20,21}, which  describes light, heavy-light mesons, and heavy
quarkonia in a universal way, only via such fundamental parameters
as string tension and the pole (current) mass of the $c$ quark.
For low-lying states it is also important to fix the value of the
vector strong coupling at large distances -- $\alpha_{\rm crit}$
(the freezing constant), but for high excitations different choice
of $\alpha_{\rm crit}$ gives small uncertainty in their  masses,
$\sim 20$~MeV.

At this point we would like to underline that widely used spinless
Salpeter equation (SSE) appears to be a particular case of RSH
with the only restriction. If in constituent potential models the
$c-$ quark mass is taken as a fitting parameter, in our approach
in SSE the $c$-quark mass has to be equal to its pole mass. At
present the pole mass of the $c$ quark is defined with a good
accuracy: $m=m_c({\rm pole})=1.40\pm 0.07$~GeV \cite{ 2}. It is of
interest that the masses of higher charmonium states appear to be
very sensitive to accepted value of $m(\rm pole)$. We also show
that if GE potential is considered as a perturbation, then the
masses $\tilde{M}(nl)~(n\geq 3)$ coincide with exact solutions of
RSH (or SSE) with an accuracy $\simeq 2\%$.

Moreover, in ``linear" approximation the masses are shown to be
defined by simple analytical expressions.

We do not consider here non-conventional charmonium resonances, in
particular, those  which occur near thresholds, since they may be
calculated only within two-(many-)channel approach \cite{16,17}.

\maketitle

\section{The masses $M(nS)$ and $M(nD)$}

Although RSH was derived for an arbitrary $q_1\bar q_2$ meson
\cite{ 19,20},  in case of heavy quarkonia it has more simple
form, because so-called string and self-energy corrections are
small and can be neglected \cite{21}:
\begin{equation}
H=\omega(nl) +\frac{m^2}{\omega(nl)}+\frac{\vep^2}{\omega(nl)}+ V_
{\rm B}(r), ~~H\varphi_{nl}=M(nl)\varphi_{nl}. \label{eq.01}
\end{equation}
We use here einbein approximation (EA) \cite{ 20,21}, when the
mass $M(nl)\equiv M_{\rm cog}(nl)$ is defined as

\begin{equation}
M(nl)=\omega(nl)+\frac{m^2}{\omega(nl)} + E_{nl}(\omega(nl)).
\label{eq.02}
\end{equation}

This mass formula does not contain any overall (fitting) constant
and depends on the pole mass of the $c$ quark $m$, which is
defined via the current mass of the $c$ quark and now known with
an accuracy $\sim 70$ MeV \cite{ 2}; in our paper we take
$m=1.40$~GeV.

In (\ref{eq.02}) a variable $\omega(nl)$ is the averaged kinetic
energy of the $c$ quark for a given $nl$ state, which plays a role
of a constituent quark mass, being  different for different
states:
\begin{equation}
\omega(nl)=\langle\sqrt{\vep^2+m^2}\rangle_{nl}. \label{eq.03}
\end{equation}
In (\ref{eq.02}) $E_{nl}(\omega(nl))$ is the excitation energy of
a given state $nl$; its depends on  static potential used. Here we
take ``linear + GE" potential $V_{\rm B}(r)$ as in \cite{12,21},
\begin{equation}
V_{\rm B}(r)=\sigma r - \frac{4\alpha_{\rm B}(r)}{3 r}.
\label{eq.04}
\end{equation}
For low-lying states both linear and GE terms are important and to
calculate $E_{nl}$, $\omega(nl)$ one needs to solve two equations
in consistent way: firstly, the equation (\ref{eq.01}) and also
the equation for $\omega(nl)$:
\begin{equation}
\omega(nl)^2=m^2+\omega(nl)^2 \frac{\partial E_{nl}}{\partial
\omega(nl)}
\label{eq.05}
\end{equation}
For higher states confining potential dominates and due to this
fact exact solutions of (\ref{eq.01}), (\ref{eq.05}) and the
masses $\tilde{M}(nl)$,  calculated for linear potential  with GE
potential taken as a correction, coincide with an accuracy better
$2\%$ (see Tables II, III).

In ``linear" approximation (with only linear potential) the
excitation energy $E_0(nl)$ is given by the expression:
\begin{equation}
 E_0(nl)=\left (\frac{\sigma^2}{\omega_0(nl)}\right)^{1/3}\zeta_{nl},
\label{eq.06}
\end{equation}
while from (\ref{eq.05}) the equation for $\omega_0(nl)$ is
\begin{equation}
{\omega_0(nl)}^2=m^2+ \frac13 (\sigma\omega_0(nl))^{2/3}\zeta_{nl}.
\label{eq.07}
\end{equation}

From (\ref{eq.06}) and (\ref{eq.07}) one can see that  $E_0(nl)$
and $\omega_0(nl)$ are expressed  via the string tension $\sigma$
and the Airy numbers $\zeta_{nl}$. It is also important that
$\omega_0(nl)$ depends on the $c-$ quark pole mass, being
proportional $m$. Through our paper the conventional values
$\sigma=0.18$~GeV$^2$ and $m\equiv m_c(\rm pole)=1.40$~GeV are
taken. The Airy numbers for $n=1,...,8~(l=0.2)$ are given in
Appendix.

The equation (\ref{eq.07}) (with $m\neq 0$) easily reduces to the
Cardano equation, from which  $\omega_0(nl)$ is  obtained in
analytical form:
\begin{equation}
\omega_0^{2/3}(nl)=\left(\frac{m^2}{2}\right)^{1/3}
 \left\{\left(1+\sqrt{1-\left(\frac{2\sigma}{27 m^2}\right)^2
\zeta_{nl}^3}\right)^{1/3}+\left(1-\sqrt{1-\left(\frac{2\sigma}{27
 m^2}\right)^2 \zeta_{nl}^3}\right)^{1/3}\right\}
 \label{eq.08}
\end{equation}
From this equation it follows  that
\begin{equation}
\omega_0=\frac{m}{\sqrt
2}\left\{\left(1+\sqrt{1-\left(\frac{2\sigma}{27 m^2}\right)^2
\zeta_{nl}^3}\right)^{1/3}+\left(1-\sqrt{1-\left(\frac{2\sigma}{27
 m^2}\right)^2 \zeta_{nl}^3}\right)^{1/3}\right\}^{3/2}. \label{eq.09}
\end{equation}

In linear approximation the  kinetic energies $\omega_0(nl)$ have
several characteristic features (see Table 1):

\begin{enumerate}
\item They differ for the states with different quantum numbers
$nl$, increasing for larger radial excitations: from 1.73~GeV for
the $4S$ state to $\omega_0(7S)=1.94$~GeV.

\item For $n\geq 3$ $\omega_0(nD)$  and $\omega_0((n+1)S)$ almost
coincide and due to this property the masses of these states are
degenerated for linear potential -- a difference between them is
$\leq 5$~MeV.

\item The masses $\omega_0(nl)$ are proportional to the $c$-quark
pole mass.

\item The values of $\omega_0(nl)$ do not practically depend on GE
interaction, coinciding with exact $\omega(nl)$ for $n\geq 3$ with
an accuracy better $3\%$ (see Table VI in Appendix).

\end{enumerate}
A growth of $\omega_0(nl)$ for larger n is an important feature of
``a constituent" mass in relativistic string approach. Due to this
property, the r.m.s. of higher charmonium states are not very
large, changing from 1.4 fm for the $4S$ state to 2.0 fm for the
$7S$ state (these radii are given in Appendix). Therefore one can
expect that higher charmonium resonances exist and can manifest
themselves in different $e^+e^-$ processes, if their leptonic widths are not small.\\

\begin{table}
\caption{ The kinetic energies $\omega_0(nl)$ $(l=0,2)$ (in GeV) from
(\ref{eq.08}) for linear potential with \\ $\sigma=0.18$~GeV$^2$
($m=1.40$~GeV).}
\bigskip
\begin{tabular}{|c|c|c|c|}
\hline
 ~~~~~~~~~ $nS$~~~~~~~~~ & ~~~~~~~ $\omega_0(nS)$~~~~~~~~ & ~~~~~~$nD$~~~~~~&~~~~~ $\omega_0(nD)$~~~~~~~~~~~\\\hline

  1S   &    1.512  &       -   &   -\\
  2S  &     1.598  &       1D  &   1.606\\
  3S  &     1.669 &        2D  &   1.674\\
  4S &      1.732 &        3D  &   1.736\\
  5S &      1.789 &        4D  &   1.793\\
  6S &      1.884 &        5D  &   1.847\\
  7S  &     1.896  &       6D  &   1.898\\
  8S  &     1.945 &        7D &    1.947\\

\hline
\end{tabular}
\end{table}

In Appendix (Table VI) the values of $\omega_0(nl)$ are compared
to ``exact" $\omega(nl)$ calculated for SSE :

\begin{equation}
\{2\sqrt{\vep^2+m^2}+V_{\rm B}(r)\} \varphi_{nl}=M(nl)\varphi_{nl},
\label{eq.10}
\end{equation}
with the same ``linear+GE" potential (\ref{eq.04}); their values
coincide with an accuracy better $3\%$, i.e. for higher
excitations $\omega(nl)$ appears to be independent of GE potential
used.

The SSE (\ref{eq.10}) may be considered as a particular case of
the RSH, in which a string correction is neglected as in
(\ref{eq.01}). It can be derived from RSH, if the extremum
condition is put as  $\frac{\partial H}{\partial \omega}=0$
\cite{22}. On the other hand, EA follows from RSH, if the extremum
condition is put on the mass (\ref{eq.02}) as $\frac{\partial
M(nl)}{\partial \omega}=0$ \cite{ 20, 22}. Here we mostly use EA,
because in this approach the wave functions (w.f.) with $l=0$ are
finite near the origin, while for SSE the $S-$wave solutions
diverge.

In Tables II, III ``exact" solutions of SSE, denoted as $M(nl)$,
are compared to approximate masses $\tilde{M}(nl)$:
\begin{equation}
\tilde{M}(nl)=M_0(nl) +<V_{GE}>_{nl},  \label{eq.11}
\end{equation}
where $M_0(nl)$ is a solution of (\ref{eq.01}) with only linear
potential. The masses $M_0(nl)$ for $l=0,2$ and the matrix
elements (m.e.) $<V_{GE}>_{nl}$ are given in Appendix, Tables VII
and VIII. The GE contribution to $\tilde{M}(nl)$ is negative, with
much smaller magnitude ($\sim 150$~MeV) than that for linear
potential, which is  $\sim 1.5-2.0$~GeV. However, this GE
correction, $\sim 10\%$, is important to improve an agreement
with known experimental masses. \\

\begin{table}
\caption{ The masses $\tilde{M}(nS)$ (\ref{eq.11}) and exact
solutions $M(nS)$ for SSE (in GeV) ($\sigma=0.18~GeV^2$,
$\alpha_{\rm crit}=0.54$) }
\bigskip
\begin{tabular}{|c|c|c|c|}
\hline

   ~~~~~~ State ~~~~~~&   ~~~~~~~  $M(nS)$ for SSE ~~~~~~& ~~~~~~~   $\tilde{M}(nl)$ ~~~~~~~~&
   ~~~~~~~experiment~~~~~~~\\

     -      &     m=1.41~GeV   &     m=1.40~GeV   &     -\\\hline

    1S       &   3.07  &            3.068  &          3.067\\

     2S   &      3.67    &           3.663    &        3.67(4)\\

     3S  &        4.09  &             4.099&              4.040\\

     4S   &       4.45      &        4.464    &          4.421\\

     5S   &        4.75    &          4.792     &         4.78$^a $ \\

     6S    &      5.04  &             5.087   &           5.09$^a$\\

     7S  &        5.31    &           5.365   &            5.44$^a$\\

      8S  &       -     &             5.630  &              5.91$^a$\\

\hline
\end{tabular}
\bigskip

  $^a$ This number is taken from the fit to experimental data \cite{
10}.

\end{table}
From Tables II, III one can see that the differences between
$M(nl)$ for SSE and $\tilde{M}(nl)$, $\simeq 40$~MeV, lie within
accuracy of our calculations. Thus our calculations (in
single-channel approximation) show that

\begin{enumerate}

\item The value $M(3S)=4.09$~GeV is by 50~MeV larger than
experimental number, $M(\psi(4040)=4.04$~GeV, and this difference
between them agrees with the value of hadronic shift for this
resonance, $\sim 40$~MeV, predicted in \cite{16}.

\item For $\psi(4415)$ a smaller hadronic shift, $\sim 30$~MeV,
follows.

\item Calculated $M(5S)=4.79$~GeV agrees with the prediction of
$M(\psi(5S))=4.78-4.82$~GeV from \cite{ 10}, \cite{ 11}, where
this mass has been extracted from fits to experimental cross
sections for different $e^+e^-$ processes \cite{6,7,8,9}. Such a
coincidence takes also place for the $M(6S)=5.09$~GeV.

\item On the contrary, our masses for the $7S,~8S$ charmonium
states: $M(7S)= 5.365$~GeV and $M(8S)=5.63$~GeV, are by 80~MeV and
$\sim 300$~MeV smaller than those from \cite{10} (see Table II). ,

\item For the $D-$wave states $M(3D)=4.54$~GeV and
$M(4D)=4.86$~GeV are obtained; the value of $M(4D)$ agrees with
$M(4D)\sim 4.87$~GeV from  \cite{ 10} (see Table III). For higher
$5D$ and $6D$ our values are smaller, by 160~MeV and 250~MeV,
respectively, than in \cite{10}.

\item For purely linear potential the spacings
$\delta_{n+1,n}=M_0(nD)-M_0((n+1)S)$ are small, $\sim 15\pm 5$~MeV
(see Tables VII, VIII), i.e., these levels are degenerated.
However, due to GE potential these mass differences increase, so
that $\tilde{M}(3D)-\tilde{M}(4S)=80$~MeV and
$\tilde{M}(6D)-\tilde{M}(7S)=50$~MeV.

\end{enumerate}

\begin{table}
\caption{ The masses $M(nD)$ for SSE and $\tilde{M}(nl)$
(\ref{eq.11}) (in GeV) with $\sigma=0.18~$ GeV$^2$ }
\bigskip
\begin{tabular}{|c|c|c|c|}
\hline

  ~~~~~~   $nD$~~~~~~~~~ &  ~~~~~~~  $M(nD)$ for SSE ~~~~~~~~~&~~~~~~~  $\tilde{M}(nD)$ ~~~~~~~&~~~~~~~~~
  experiment~~~~~\\
              &   $m=1.41$~GeV  &      $m=1.40$~GeV&\\\hline

       1D   &        3.80  &           3.80  &        3.77\\
       2D  &         4.18  &           4.192    &     4.16\\
       3D   &        4.51    &        4.543    &     4.55$^a$\\
       4D    &       4.81  &          4.854     &   4.87$^a$\\
       5D    &       5.09  &          5.143      &   5.30$^a$\\
       6D    &       5.35 &           5.413    &    5.66$^a$\\
       7D    &       5.62 &           5.669    &     -\\\hline

\end{tabular}
\bigskip

$^a$ See the footnote to Table 2.
\end{table}

Here in our analysis of high charmonium excitations  we do not use
flattening potential, introduced for light mesons to take
indirectly into account a creation of virtual $q\bar q$ pairs ($q$
is a light quark) \cite{18}. Such  flattening of confining
potential was useful for light mesons, which have large hadronic
shifts. The situation in charmonium is supposed to be different,
because for higher states the $c$-quark kinetic energy increases,
being $\sim 1.7-1.9$~GeV, and one can expect that their hadronic
shifts are not large ($\leq 40$~MeV) and their overlapping
integrals, which describe different decay modes, are smaller than
those for low-lying resonances.

In \cite{23} the masses of higher charmonium states have been
calculated with the use of a static potential, which contains a
large number of additional parameters and large overall constant,
while the value of the string tension is relatively small.
Nevertheless calculated in \cite{23} masses of the $nS~(n=5,6)$
and $nD~(n=3,4,5)$ charmonium states coincide with our predictions
within $\pm 50$~MeV, while in \cite{23} $M(\psi(6D)=6.03$~GeV is
by 260~MeV larger than in our calculations.

At this point we would like to stress that with the use of RSH all
calculated masses do not contain a fitting constant and totally
defined only by $\sigma=0.18$~GeV$^2$, $m(\rm pole)=1.40$~GeV,
while a choice of the freezing value of the strong coupling
$\alpha_{\rm crit}$ is not very important.

\section{Radial Regge trajectories for the $nS$ and $nD$ states}

The Regge trajectories, orbital and radial, are usually studied in
light mesons and now it remains unclear whether a regime of linear
trajectories takes place for the charmonium family or not. In
\cite{24} it was assumed that linear Regge trajectories describe
charmonium states with different quantum numbers  with an accuracy
$\sim 100$~MeV, while the slopes were defined fitting the masses
of low-lying (well-established) charmonium states.

Here from our dynamical calculations of the $M(nS)$ and $M(nD)$ it
follows that linear Regge trajectories take place only for higher
charmonium states.

The radial Regge trajectory can be presented as:
\begin{equation}
M^2(nl)= \mu_l^2 +\Omega_l~ n, \label{eq.12}
\end{equation}
where $\mu_l$ and the slope $\Omega_l $ are supposed to be
constants. In classical string picture for massless quarks
$\Omega_l=4\pi\sigma=2.26$~GeV$^2$ ($\sigma=0.18$~GeV$^2$),
however, for light mesons the values of $\Omega_l$ have appeared
to be smaller, $ 1.3-1.6$~GeV$^2$, because of large hadronic
shifts \cite{18}.

Here we consider the masses of the centers of gravity and define
the Regge trajectories for a given $l$, when from (\ref{eq.12})
the spacing between squared masses:
\begin{equation}
\Delta_{n+1,n}=\tilde{M}^2((n+1)l)-\tilde{M}^2(nl)=\Omega_l
\label{eq.13}
\end{equation}
has to be  a constant $\Omega_l$. Taking from \cite{2}
experimental values of the c.o.g. masses for $J/\psi-\eta_c(1S),~
\psi(3686)-\eta_c(2S),~ \psi(4040),~ \psi(4415)$  one obtains that
the spacing $\Delta_{21}=4.07$~GeV$^2$, while
$\Delta_{32}=2.84$~GeV$^2$ is significantly smaller, and
$\Delta_{43}=3.22$~GeV$^2$ is by $\sim 15\%$ larger than
$\Delta_{32}$. A decrease of $\Delta_{32}$ possibly occurs due to
hadronic shift of the $\psi(4040)$ resonance, which is $\sim
50$~MeV. If one takes unshifted masses from Table II:
$M(3S)=4.099$~GeV and $M(4S)=4.464$~GeV, then
$\Delta_{32}=3.32$~GeV$^2$ and $\Delta_{43}=3.13$~GeV$^2$ become
close to each other, still being larger than $\Omega_S$ for higher
states (calculated $\Delta_{n+1,n}$ are given in Table IV).

\begin{table}
\caption{  The differences $\Delta_{(n+1),n}$ (in GeV$^2$) between
squared masses $\tilde{M}^2(nl)$ for the $nS$ states}
\bigskip
\begin{tabular}{|c|c|}
\hline
~~~~~~~~~~~ $\Delta_{43}$ ~~~~~~~~~~~~~~& ~~~~~~~~~~~~3.22~~~~~~~~~~~\\

  $\Delta_{54}$   & 3.04\\

  $\Delta_{65}$   &  2.91\\

  $\Delta_{76}$   &   2.91\\

   $\Delta_{87}$    & 2.91\\\hline

\end{tabular}

\end{table}

The numbers from Table 4 show that $M(nS)$ with $4\leq n\leq 8$
can be described by linear (radial) Regge trajectory with the
slope

\begin{equation}
\Omega_S = 2.91 ~{\rm GeV}^2, \label{eq.14}
\end{equation}
which is a constant with a good accuracy. From here

\begin{equation}
M^2(nS)=M^2(4.21)+2.91~{\rm GeV}^2 (n-4). \label{eq.15}
\end{equation}

For the masses $M(nD)$ the slope $\Omega_D$ slightly decreases
changing from $\Delta_{43}=3.13 $~GeV$^2$ to a smaller value,
$\Delta_{76}=2.84$~GeV$^2$ (see masses from Table III). Therefore
for the $nD$ states their masses are described by linear Regge
trajectory with worse accuracy than for the  $nS$ excitations,
giving

\begin{equation}
\Omega_D=(2.88\pm 0.04)~GeV^2, \label{eq.16}
\end{equation}
and
\begin{equation}
M^2(nD)=M^2(4.54)+\Omega_D~(n-3). \label{eq.17}
\end{equation}

In \cite{24} charmonium states with different quantum numbers,
including low-lying states, were described by linear Regge
trajectories with $\mu_l^2$ and the slopes $\Omega_J$, defined
from fits to known experimental masses. For the masses of
$\psi(nS)$ and $\psi(nD)$ the slope
$\Omega_S=\Omega_D=3.2$~GeV$^2$ was obtained, which is only 10\%
larger than that in our dynamical calculations, while the values
of $\mu_S=2.6$~GeV and  $\mu_D=3.31$~GeV in \cite{24} are taken as
fitting parameters. In our calculations linear Regge trajectories
can be applied only to higher charmonium states and
$\mu_l~(l=0,2)$ is equal to experimental mass.

Moreover, for the slopes $\Omega_S, \Omega_D$ approximate
analytical expressions can easily be derived. If in (\ref{eq.02}),
(\ref{eq.06}) one takes an averaged $\bar \omega_{0}=\bar
\omega_S=\bar \omega_D$ for a kinetic energies with $n\geq 4$,
then the slope
\begin{equation}
\Omega_l =\left(\frac{\sigma^2}{\bar \omega_{0}}\right)^{1/3}
(\zeta_{(n+1)l}-\zeta_{nl}) \left\{ \left( \frac{\sigma^2}{\bar
\omega_{0}}\right)^{1/3}(\zeta_{(n+1)l}+\zeta_{nl}) +2 \bar
\omega_{0}+\frac{2m^2}{\bar \omega_{0}}\right\}
\label{eq.18}
\ee
is fully defined by $\bar \omega_0$ and the Airy numbers. From
(\ref{eq.18}), taking $\bar \omega_{0}\simeq 1.84$ GeV
($\zeta_{nS}, \zeta_{nD}$ are given in Table V), one obtains
$\Omega_S\simeq \Omega_D\simeq 2.90$~GeV$^2$ in good agreement
with ``exact" number in (\ref{eq.14}), (\ref{eq.16}).

It is important to stress that in charmonium the slopes
$\Omega_S, \Omega_D$ have appeared to be  two times larger than
those for light mesons \cite{18}.

For calculated masses a spacing between neighbouring radial
excitations, $M((n+1)l)-M(nl),$ is large, being $\sim 300$~MeV for
the $nS$ states and $\sim 270$~MeV for the $nD$ states. On the
contrary, the mass difference $M(nD)-M((n+1)S)$ is smaller,
decreasing from $\sim 120$~MeV for low-lying states to $\sim
80-60$~MeV for large $n$. Evidently, that for such small spacings
the $S-D$ mixing has to be important. Then the $S-D$ mixing
strongly affects dielectron widths of vector charmonium states. As
shown in \cite{21}, higher $nS$ states have large dielectron
widths, $\sim 1$~KeV, which are by two orders larger than
$\Gamma_{ee}(nD)$ for purely $D-$wave states, e.g.
$\Gamma_{ee}(1D)\simeq 15$~eV and $\sim 40$~ eV for the $4D$
state. Therefore purely $nD$ resonances with such small dielectron
widths cannot be observed in the $e^+e^-$ experiments, while they
may be seen, if due to the $S-D$ mixing, their dielectron widths
are of the same order as those of the $nS$ states.
\section{Conclusions}

Our calculations of higher charmonium states were performed in
single-channel approximation when the $c\bar c$ dynamics at large
distances can be studied in detail. With the use of RSH we have
obtained that

\begin{enumerate}

\item  The $5S-8S$ and $3D-7D$ states occur in the range 4.5-5.8
GeV and the spacing between neighbouring radial excitations is of
the order of 250-300~MeV for $n\geq 4$.

\item The mass differences between $M(nD)$ and $M((n+1)S)$ are
rather small, decreasing from $\sim 80$~MeV for $n=4$ to $\sim
50$~MeV for $n=7$. The important point is that for purely linear
potential these levels are degenerated (their mass difference is
$\simeq 15$~MeV), while due to GE interaction a spacing
$M(nD)-M((n+1)S$ increases.

\item The masses of radial excitations,  $M(nS)$ and $M(nD)$ with
$n\geq 3$, are described with good accuracy by linear Regge
trajectories with the slope $\Omega_S=2.91$~GeV$^2  $ and
$\Omega_D=2.88$~GeV$^2$.

\item  The masses of high excitations in charmonium are mostly
defined by linear confining potential and at the same time they
depend on the pole mass of a $c-$quark. Here $m(\rm
pole)=1.40$~GeV is used.

\item Higher $nD$ resonances can be observed in experiments only
if their dielectron widths are of the same order as those for the
$nS$ states, which happens due to the $S-D$ mixing.

\item We predict the following masses: $M(3D)=4.54$~GeV, $M(5S)=4.79$~GeV,\\
$M(4D)=4.85$~GeV, $M(6S)=5.09$~GeV, $M(5D)= 5.14$~GeV,
$M(7S)=5.365$~GeV, $M(6D)=5.41$~GeV, $M(8S)=5.63$~GeV, and
$M(7D)=5.67$~GeV. An accuracy of our calculations is estimated to
be $\sim 20$~MeV, if hadronic shifts are neglected.

\end{enumerate}
These characteristic features of the $c\bar c$ dynamics at large
distances can be tested by future experiments in which the masses
and dielectron widths of higher charmonium resonances have to be
measured with precision accuracy.

Acknowledgements.

I am grateful to Prof. Yu.A.Simonov for useful discussions. This
work was supported by the RFFI Grant 09-02-00629a.

\newpage

\setcounter{equation}{0}
\renewcommand{\theequation}{A.\arabic{equation}}

\hfill {\it  Appendix  }

\centerline{\bf The  matrix elements, $M_0(nl)$, and the Airy
numbers}

 \vspace{1cm}

Firstly, we give the Airy numbers for the $nS$ and $nD$ states.\\

\begin{table}
[h]
\caption{  The Airy numbers $\zeta_{nl}$ for $l=0,2$.
 }
\bigskip
\begin{tabular}{|c|c|c|c|}
\hline ~~~~~~~~~~~
   $nS$ state  ~~~~~&  ~~~~~~ $\zeta_{nS}$~~~~~~&~~~~~~~   $nD$ state~~~~~&~~~~~~~
   $\zeta_{nD}$~~~~~\\ \hline

      1S   &       2.338107   &    -  &            -\\
      2S  &        4.087949  &     1D &         4.24818\\
      3S  &        5.520560  &     2D    &      5.62971\\
      4S  &        6.786708   &    3D   &       6.86889\\
      5S  &        7.944134 &      4D   &       8.00981\\
      6S &         9.022651   &    5D &         9.07700\\
      7S&        10.040174  &      6D   &       10.08646\\
      8S  &        11.008524 &       7D &         11.04874\\
\hline

\end{tabular}

\end{table}
Using the Airy numbers, one can calculate the kinetic energies
$\omega_0(nl)$ (\ref{eq.09}) as well as excitation energies
$E_0(nS)$ and $E_0(nD)$ (\ref{eq.06}) for purely linear potential.
In Table VI $\omega_0(nS)$ and $\omega_0(nD)$ for linear potentil
and also ``exact" $\omega(nS)$, calculated for SSE, are given.
\begin{table}

\caption{  The averaged kinetic energies $\omega_0(nS)$, $\omega_0(nD)$ (in
GeV) for linear potential and $\omega(nl)$ from SSE ($\sigma=0.18$~GeV$^2$).
 }
\bigskip
\begin{tabular}{|c|c|c|c|c|}
\hline ~~~~~  state
~~~&~~~~$\omega_0(nS)$~~~&~~$\omega(nS)$~~~&~~state~~~~&~~~~~~~
  $\omega_0(nD)$~~~~~\\\hline
    1S  &   1.512   &         1.60   &          -  &    -\\
    2S  &   1.598    &        1.66    &         1D  &  1.606\\
    3S  &   1.669    &        1.73    &        2D   &  1.674\\
    4S  &   1.732    &        1.78    &         3D  &  1.736\\
    5S   &  1.789    &        1.84    &         4D   & 1.793\\
    6S   &  1.844    &        1.88     &        5D   &  1.847 \\
   7S    &  1.896     &      1.94  &             6D  &  1.898\\
    8S   &   1.9449  &        -      &           7D  & 1.9470\\
\hline

\end{tabular}

\end{table}

As seen from Table VI, for linear potential the kinetic energies
$\omega_0((n+1)S)$ and $\omega_0(nD)$ practically coincide for all
$n$, while ``exact" $\omega(nS)$, calculated for SSE, differ from
$\omega_0(nS)$ only by $\leq 3\%$. It means that $\omega(nS)$
weakly depends on GE potential taken and only for low-lying states
a difference between them is $\sim 6\%$.

Knowing $\omega_0(nl)$ one can define the excitation energy
$E_0(nl)$, the total mass $\tilde{M}(nl)$, and also the w.f. at
the origin  for a given state $nl$. The excitation energies
$E_0(nS)$ and $E_0(nD)$ are given in Tables VII, VIII together
with r.m.s. $\sqrt <r^2>_{nl}$ and m.e.
$<V_{GE}(r)>_{nl}$.\\

\begin{table}
 [b]
\caption{  The values $E_0(nS)$, $M_0(nS)$, the m.e.
$<V_{GE}(r)>_{nS}$ (in GeV), and  $\sqrt<r^2>_{nS}$ (in fm) for
linear potential with $\sigma=0.18$~GeV$^2$, $m=1.40$~GeV.
 }
\bigskip
\begin{tabular}{|c|c|c|c|c|}
\hline ~~~~~~~~~~~

 nS state~~~~~&~~~~~  $E_0(nS)$~~~~~&~~~~~ $ M_0(nS)$~~~~~&~~~~~$ <V_{GE}(r)>_{nS}$
 ~~~~~&~~~~~$\sqrt<r^2>_{nS}$\\\hline

   1S  &      0.6494  &    3.458   &   - 0.390    &    0.519\\
   2S    &    1.1147  &    3.939   &   - 0.276    &    0.891\\
   3S    &    1.4837  &    4.327   &   - 0.228  &      1.186\\
   4S     &   1.8016  &    4.665   &    -0.201   &     1.440\\
   5S   &     2.0861 &     4.971   &    -0.179  &      1.667\\
   6S    &    2.3454   &   5.252   &    -0.165  &      1.875\\
   7S   &     2.5861  &    5.516   &    -0.151   &     2.067\\
   8S   &     2.8115   &    5.764  &     -0.134   &     2.250\\
\hline

\end{tabular}

\end{table}
For the $D-$wave states a contribution from GE potential is
smaller (see $<V_{GE}(\rm r)>_{nD}$ in Tables VI, VII);  due to
this fact the mass differences $\tilde{M}(nD)-\tilde{M}((n+1)S)$
increase.

\begin{table}
\caption{  The  values $E_0(nD)$, $M_0(nD)$, the m.e.
$<V_{GE}(r)>_{nD}$ (in GeV), and $\sqrt<r^2>_{nD}$ (in fm) for
linear potential with $\sigma=0.18$~GeV$^2$, $m=1.40$~GeV.
 }
\bigskip
\begin{tabular}{|c|c|c|c|c|}
\hline ~~~~~~~~~~~

 nD state~~~~~&~~~~~  $E_0(nD)$~~~~~&~~~~~ $ M_0(nD)$~~~~~&~~~~~$ <V_{GE}(r)>_{nD}$
 ~~~~~&~~~~~$\sqrt<r^2>_{nD}$\\\hline

   1D   &         1.5645   &        3.983   &      -0.183    &         0.879 \\
   2D     &       1.5114     &      4.356     &   - 0.164      &       1.179 \\
   3D     &       1.8219     &      4.687     &   - 0.144      &       1.436 \\
   4D      &      2.1017      &     4.988      &  - 0.134       &      1.664 \\
   5D    &        2.3584    &       5.267    &    - 0.124     &        1.872 \\
   6D     &       2.5970     &      5.528     &   - 0.115      &       2.065 \\
   7D    &        2.8208    &       5.775    &    - 0.106     &        2.246 \\
   \hline

\end{tabular}

\end{table}
In our calculations of $\tilde{M}(nl)=M_0(nl)+<V_{GE}>_{nl}$ for
higher states the strong coupling  $\alpha_B(r)=\alpha_{\rm
crit}=\rm constant$ was taken, i.e., in the GE potential
$V_{GE}=-\frac{4\alpha_{\rm crit}}{3r}$ the asymptotic freedom
behavior of the strong coupling  was neglected , since it gives
negligible correction for high excitations. Here the value
$\alpha_{\rm crit}=0.54$ and $<V_{GE}>_{nl}=-0.72<r^{-1}>_{nl}$
are used.

\end{document}